\documentclass[aps,prl,twocolumn,groupedaddress]{revtex4}
\usepackage{amsmath}
\usepackage{amssymb}
\usepackage{graphicx}

\newcommand{\R}{\mathbb{R}}
\newcommand{\ca}[1]{{\mathcal #1}}
\newcommand{\Rx}[3]{{{\cal R}_{#1,#2}(#3)}}
\newcommand{\mnorm}[1] {\bigl\Vert #1 \bigr\Vert}
\newcommand{\snorm}[1] {\Vert #1 \Vert}
\newcommand{\RxB}[2]{{{\cal R}_{#1,#2}^*}}
\newcommand{\E}{\mathbb{E}}  
\newcommand{\Lx}[2]{L_{#1} (#2)}

\def \e         { \varepsilon }
\def \vt        { \vartheta }
\def \s         { \sigma }
\def \lb        { \lambda }
\def \g         { \gamma }
\def \om        { \omega }
\def \p         { \varphi }

\DeclareMathOperator{\corpo}{cor}

\begin{document}

\title{ Forecasting the Evolution of Dynamical Systems from Noisy Observations }

\author{M.\ Anghel and I.\ Steinwart }

\affiliation{Los Alamos National Laboratory, Los Alamos,
NM 87545}
\email[]{{manghel,ingo}@lanl.gov}
\date{\today}

\begin{abstract}

We consider the problem of designing almost optimal predictors for
dynamical systems from a finite sequence of noisy observations and
incomplete knowledge of the dynamics and the noise.  We first discuss the
properties of the optimal (Bayes) predictor and present the
limitations of memory-free forecasting methods, and of any finite memory
methods in general.  We then show that a nonparametric support vector
machine approach to forecasting can consistently learn the optimal
predictor for {\em all} pairs of dynamical systems and bounded
observational noise processes that possess summable correlation
sequences. Numerical experiments show that this approach adapts the
memory length of the forecaster to the complexity of the learning task
and the size of the observation sequence.

\end{abstract}

\maketitle

Our goal is to design almost optimal predictors for dynamical systems
conditioned only on a finite sequence of observed {\em noisy}
measurements of its evolution and {\em imperfect} knowledge of the
dynamics and noise.  Suppose that $F:X\to X$ is a map on a compact subset
$X\subset \R^d$ such that the associated dynamical system $\ca D :=
(F^i)_{i\geq 0}$, where $F^i=F \circ F^{i-1}$, has an ergodic measure $\mu$.
For example, if $X$ is compact and $F$ is continuous such a measure
always exists.  Moreover, let $\ca E := (\e_i)_{i\geq 0}$ be an
$\R^d$-valued i.i.d.~process with respect to the distribution $\nu$
that is independent of the process $\ca D$.  We assume that all
observations of the dynamical system are perturbed by the process $\ca
E$, i.e.  all observations are of the form 
\begin{equation}\label{obs_process}
z_i:= F^i(x_0) + \e_i\, ,
\end{equation}
where $x_0$ is the unknown 
initial point of the trajectory.  Let us now assume
that we have a sequence of observations $z_1,\dots,z_{m}$ and we wish
to predict the observation $z_{m+l}$, or the true state $x_{m+l}$,
after some time $l$ as best as possible.  Then one possible
formalization of this problem is to look for a minimizer $f^*_{m,l}$,
called the Bayes forecaster, of the risk functional 
\begin{align}\label{opt-fore-prob}
&\Rx mlf = \int_{\R^{d}}\int_{\R^{(m+1)d}} \mnorm{ F^{m +l}(x) + \e_{m+l} \\
&-  f   \bigl( F(x)+ \e_1,\dots,F^{m}(x)+\e_{m}   \bigr)   }_p^p\,  \nu^{m+1}(d\e)\mu(dx)\, , \notag
\end{align}
which describes the average discrepancy (measured in the 
$\snorm \cdot _p$-norm) between the predictions of $f:\R^{md} \to \R^d $ 
and the observations (we have made a
time-shift by $m+1$ in order to handle non-invertible maps $F$).
To attain this goal we want to design a {\em learning} method $\cal{L}$, 
which assigns to each observed trajectory 
$T_n = \{z_0,\dots,z_n\}$, $ n \ge m$, a forecaster $f_{T_n}$ and 
which is {\em consistent}, i.\ e.\
\begin{equation}\label{consistency}
 \lim_{n \to \infty} \Rx ml{f_{T_n}} = \RxB{m}{l} 
\end{equation}
holds  in probability, where $\RxB{m}{l} = \Rx ml{f_{m,l}^*}$ is the minimum 
risk (the Bayes risk).

Most classical forecasting methods use a Markov state-space approach
to modeling dynamic systems and require a model of both the flow $F$
and the observational noise $\nu$.  These methods usually attempt to
estimate the probability density function (PDF) of the state based on
all the available information, i.~e. $p(X_m|Z_{1}^m)$, where $Z_1^m:=
(Z_1,\dots,Z_m)$ and $(Z_i)$ denotes the stochastic process
that generates the observations. Then, an optimal (with respect to any criterion)
forecaster may be obtained from the PDF, including
$\E(X_{m+l}|Z_{1}^m) = \int p(X_m|Z_{1}^m) F^l(X_m) d X_m $.
 Known as nonlinear
filters, these methods estimate {\em recursively} in time this
distribution and consist of essentially two steps: a prediction step,
that uses the system model to propagate the state PDF to the next
observation time, and an update step, that uses the latest observation
to modify the prediction PDF using Bayes' rule.
Except in a few special cases, including linear Gaussian state space
models (Kalman filter) and hidden finite-state space Markov chains,
the recursive propagation of the posterior PDF cannot be determined
analytically \cite{Ma79}.
Consequently, various approximations strategies to the optimal
solutions have been developed. The most popular algorithms, the
extended and the unscented Kalman filter, 
rely on anlytical approximations of
the flow or/and finite moment approximations of the posterior PDF.
Alternatively, sequential Monte Carlo  methods, 
have conceptually the  advantage of not being subject to the assumptions of
linearity or Gaussianity in the model, but are computationally very expensive
and suffer from the degeneracy of the algorithm~\cite{DoGoAn00}.

\begin{figure}
\includegraphics[width=3.2in, height=1.8in, angle=0]{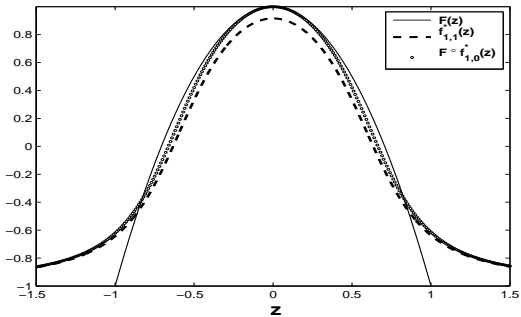}
\caption{ {\small Comparison of three, memoryless, forecasting strategies 
for  one-step ahead predictions for the logistic map and large Gaussian 
noise  ($\s=0.2$): the true dynamical behavior, $F$
(continuous line), the best forecaster (dashed), and $F$ composed
with the  best memory-free denoiser, $f^*_{1,0}$ (dotted).  }  }
\label{one_step_forecasters}
\end{figure}

Notwithstanding their significant merits, there are many difficulties
in these approaches to forecasting.  First, due to their reliance on
system and noise models they are sensitive to model errors. Second,
instead of solving the function estimation problem directly, for which
the available information might suffice, they estimate densities as a
first step, which is a harder and more general problem that requires
a large number of observations to be solved well. Third, by relying
on a Markov state-space approach to forecasting they heavily restrict
the class of functions available for forecasting.
In order to further appreciate this aspect, and to better understand
our proposed approach, let us now describe the structure of the
optimal forecasters in more detail for the special case $p=2$.
If the noise has no systematic bias, i.e. $\E_\nu \e_i = 0$, simple
algebra then shows that {\em $f^*_{m,l }$ is not only the optimal
forecaster for the observable state $z_{m+l}$ but also for the true
state $x_{m+l}$.}  Moreover, 
it is well known that 
$f^*_{m,l}= \E(Z_{m+l}|Z_{1}^m)$, but this
closed form does in general not solve the issue of actually computing
$f_{m,l}^*$ since for many cases this computation is intractable even
with perfect knowledge of $F$, $\mu$ and $\nu$.
However, if the noise has a density $\vt$
with respect to the Lebesgue measure,  $f^*_{m,l}$ can  be
expressed by a more tractable formula, 
\begin{align}\label{lstep-predictor}
&f_{m,l}^*(z_1,\ldots,z_m) = \\
&\frac{\int_{\R^d}  \vt ( z_{1}-F(x) )\cdots \vt ( z_{m}-F^{m}(x) ) F^{m+l}(x) d\mu(x)}
{\int_{\R^d} \vt ( z_{1}-F(x) )\cdots \vt ( z_{m}-F^{m}(x) )  d\mu(x)} \, . \notag
\end{align}
 When perfect knowledge of the system is available, the above equation
describes how to build the optimal predictor. Even though these
assumptions are rarely met in practice
the above result points to a number of interesting and very general
conclusions with regard to forecasting the future of a dynamical
system based on noisy observations:

{\em 1) In general the flow $F$ is not the optimal predictor.}  Indeed
Eq.~(\ref{lstep-predictor}) clearly shows that the flow $F$ is in
general not the optimal one-step predictor based only on a noisy
estimate of the present state of the system, i.e.~$F \neq
f^*_{1,1}$. A good example is provided by the logistic map, defined as
$ F(x) = 1- a x^2,$ $ x \in [-1,1]$, which is known to be ergodic and
has an invariant measure given by $ \mu(x) = 1 / \pi (1-x^2)^{1/2} $
\cite{BeSc93}.  As we can clearly see from
Fig.~(\ref{one_step_forecasters}), the true dynamical behavior $F$
(continuous line) and the best forecaster $f_{1,1}^*$ (dashed line)
disagree.  However, in the absence of observational
noise, we obviously obtain $F =
f^*_{1,1}$  from Eq.~(\ref{opt-fore-prob}), i.e.~$F$ is the optimal, memoryless, one-step ahead
predictor.

\begin{figure}[h!]
\includegraphics[width=3.2in, height=1.8in, angle=0]{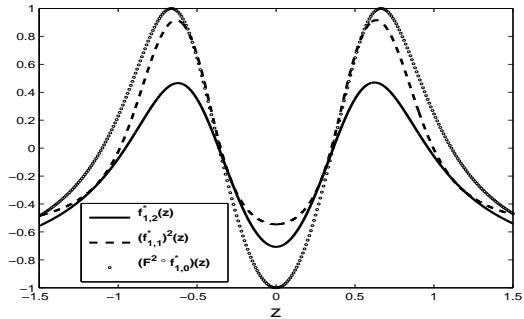}
\caption{ {\small Comparison of the direct (continuous line) and recursive
 forecaster (dashed line) for two-step ahead predictors for the
 logistic map with large Gaussian noise with $\s=0.2$. Also shown is
 $F^2$ composed with the best memory-free denoiser, $f^*_{1,0}$
 (dotted line). } }
\label{recursive-vs-direct}
\end{figure}

{\em 2) Recursive forecast is worse than direct forecast.}  Since in
general for $l$ step ahead predictors we have $f_{m,l}^* \neq
(f_{m,1}^*)^l$ we see that iterating one-step-ahead forecasts is in
general worse than directly forecasting $l$-steps ahead. In
Fig.~(\ref{recursive-vs-direct}), where we plot the direct two-step
ahead forecaster as well as the recursive forecaster obtained by
iterating the optimal one-step ahead Bayes predictor, illustrates this
effect for the logistic map. In general, as the forecasting time
increases the iterated forecasters become more and more inadequate.
Of course, for noiseless observations direct and recursive predictors
have the same forecasting performance.

{\em 3) Forecasts based on denoising the present state are not
 optimal.}  By choosing $l=0$ in Eq.~(\ref{lstep-predictor}) we obtain
 an optimal estimate ({\em optimal denoiser}) of the present state of
 the dynamical system based on a history of length $m$ \cite{rem1}.
 Now a common forecasting strategy is to estimate the present system
 state first and then apply the dynamics $F^{l}$ to this estimate.  It
 should be is clear from our discussion so far that even if the
 estimate of the present state is optimal it still approximates the
 true state: therefore, this forecaster is not the optimal $l$-step
 ahead forecaster based on histories of length $m$.  The dotted line
 in Fig.~(\ref{one_step_forecasters}), which is $F$ composed with the
 best memory-free denoiser, $f^*_{1,0}$, and the one in
 Fig.~(\ref{recursive-vs-direct}), which is $F^2$ composed with
 $f^*_{1,0}$, illustrate the principal limit of memory-free, and of
 any {\em finite} memory denoising, followed by propagating the
 dynamics.

{\em 4) Memory improves forecasting performance.}  Indeed, building
one-step ahead predictors using the past $m$ states is a minimization
of the risk functional over the space of all measurable functions
$f:\R ^{m}\to \R$. This function space is a subset of the space of
measurable functions $f:\R^{m+1} \to \R$ which is the hypothesis space
for the best one-step ahead predictor using the past $m+1$
states. Therefore, we necessarily have $ \RxB{m}{l} \geq
\RxB{m+1}{l}$, i.e.~by increasing the memory we can decrease the best
possible prediction error: building forecasters that use information
from the recent past of the dynamical system can substantially improve
the forecasting performance.  Indeed, as Table 1 shows, using
histories of increased length reduces the Bayes risk for predictors of
the logistic map.    
\begin{table}
\begin{center}
\begin{tabular}{|c|c|c|c|c|}
\hline
History  & Forecaster & $l=1$  & $l=2$   & $l=3$\\
\hline
$m=1$ & $f_{1,1}^*$ & 0.0171 & 0.0774 & 0.2082\\
 & $F \circ f_{1,0}^*$ & 0.0171 & 0.0792 & 0.2605\\
& $F$ & 0.0222 & 0.1716 & 3.8374\\
\hline
$m=2$ & $f_{2,1}^*$ & 0.0132 & 0.0543 & 0.1584\\
 & $F \circ f_{2,0}^*$ & 0.0132 & 0.0555 &  0.1868\\
\hline
\end{tabular}
\end{center}
\caption{ {\small Risks of the optimal predictor, $f_{1,1}^*$, the
denoised true dynamic, $F \circ f_{1,0}^*$, and the true dynamic $F$
for histories of length $m=1, 2$ and for the logistic map with
Gaussian noise with $\s=0.05$. } }
\end{table}

The last two features clearly show the limitations of memory-free
forecasting methods, and of any finite memory methods in general,
because they demonstrate that optimal estimators are non-Markovian in
the original state space even though the underlying dynamical system
is deterministic.  Unfortunately, it is not clear how prior knowledge
about the system can be used to build non-Markovian forecasters. For
this reason, we propose to build nonparametric predictors without
making any assumptions about the form of the dynamics or the noise,
while only assuming that the process described by (\ref{obs_process})
is bounded, ergodic, and has sufficiently fast decay of correlations.
While nonparametric methods that are proven to work in a certain
statistical sense different from (\ref{opt-fore-prob}) for arbitrary,
{\em unknown\/} stationary ergodic processes $(Z_i)$ exist
\cite{GyKoKrWa02}, these methods require very large data segments for
acceptable precision.  Approaches for forecasting goals closer to
(\ref{opt-fore-prob}) were considered by, e.g., \cite{MoMa98a,Meir00a}
but unfortunately these methods require certain mixing conditions that
cannot be satisfied by dynamical systems.

The approach we propose uses support vector machine (SVM)
forecasters~\cite{ScSm02} .  For simplicity we only describe the least
squares loss, $p=2$, and memoryless, $m=1$, one-step ahead
forecasters, $l=1$, and consider $d=1$, but generalizations are
straightforward.  An SVM forecaster assigns to each finite sequence
 $T_n$ a function $ f_{n,1,1,\lb,\g} : X \to
\R$ from a reproducing kernel Hilbert space (RKHS) $H_\g$
\cite{CuSm02a} that solves
\begin{equation}\label{svm}
\min_{f\in H_\g} \Bigl( \lb\snorm f_{H_\g}^2 + 
  \frac 1 {n} \sum_{i=0}^{n-1}  \bigl(  z_{i+1} - f( z_{i})  \bigr)^2 \Bigr) \, ,
\end{equation}
\noindent where $\lb>0$ is a free regularization parameter. Here, we
choose the RKHS of a Gaussian kernel $k_\g : X \times X \to \R$
defined by $k_\g (x,x') = \exp(-\g^2 \snorm{x-x'}_2^2 )$, where $\g >
0$ is a free parameter called the width, but other choices of kernels
are possible as well.  It can be shown that the function minimizing
the regularized empirical error (\ref{svm}) exists, is unique, and has
the form $f_{n,1,1,\lb,\g}= \sum_{i=1}^n \alpha^*_i k_\g(z_i,\cdot)$,
where $\alpha^*_1,\ldots,\alpha^*_n$ is the unique solution of the
well-posed linear system in $\R^n$
\begin{equation} \label{svm-matrix-eq}
(\lb n I + K ) \alpha^* = Z\, .
\end{equation}
\noindent Here $I$ is the $n \times n$ identity matrix, $ K$ is the $n \times n$  matrix
whose entry $(i,j)$ is $k_\g(z_i, z_j)$, where $i,j=0,1,\ldots,n-1$, and $Z$ is the  
$n \times 1$ vector $(z_1, \ldots, z_n)^T$ \cite{CuSm02a}.

Unfortunately, standard learning theory cannot make conclusions on the behavior of
$\Rx 11 { f_{n,1,1,\lb,\g}}$, since the input/output pairs
$(z_{i} ,z_{i+1}) $ are clearly not i.i.d.~samples. On the other hand,
because the observational noise process is weakly mixing and the
dynamical system is ergodic, the stochastic process $\bar{Z}_i =
(F^i,\e_i,\e_{i+1})$ defined by these pairs is ergodic. Thus, according
to Birkhoff's ergodic theorem, it satisfies the following law of large
numbers (LLN)
\begin{equation}\label{slln-l1}
\lim_{n\to \infty}\frac 1 n \sum_{i=1}^n h \circ \bar{Z}_i(\om) = \E_{\mu\otimes \nu} h\, ,
\end{equation}
$\mu\otimes \nu$-almost everywhere, for all $h \in \Lx 1 {\mu\otimes
\nu}$ defined by $h \circ \bar{Z}_i = h(z_i,z_{i+1})$.  In particular,
this holds for $h(z_i,z_{i+1})= ( z_{i+1} - f( z_{i}))^2$ for any $f
\in H_\g$. Therefore, the results in \cite{StHuSc06b} show that there
exists a null-sequence $(\lb_n)$, depending on $\ca D$ and $\ca E$,
for which the SVM forecaster is {\em consistent}.  We keep $\g$
constant, but we could as well find a sequence of the regularization
parameter {\em and} the kernel width, $(\lb_n,\g_n)$, ensuring
consistency as long as $\lim_{n\to \infty} \lb_n \g_n^d = 0$.  This
result even holds for {\em unbounded}, but integrable, i.i.d.~noise
processes.  

In order to a-priori determine for a given sample size $n$ the
regularization sequence $(\lb_n,\g_n)$,  the convergence speed of the LLN that the
observation-generating process satisfies is necessary. However, the
general negative results from \cite{Nobel99a} strongly suggest that
there exist neither a {\em universal\/}, i.e.~system and noise
independent, sequence $(\lb_n, \g_n)$ nor any other universal
forecaster.  However, the situation changes dramatically, if one
restricts considerations to dynamical systems whose stochasticity can
be described by, e.g., convergence rates of the correlations
\begin{equation}
\corpo (\psi,\p,n) := \int \psi \cdot \p\circ F^n \, d\mu - \int \psi\, d\mu\int\p\, d\mu\, 
\end{equation}
for $n\to \infty$.  
Indeed, we recently showed the existence of a {\em universal}
 sequence $(\lb_n, \g_n)$ that yields an SVM which is
consistent  for {\em all} pairs of  ergodic dynamical
 systems and {\em bounded}, i.i.d~observational noise processes satisfying
 $\sum_{n=0}^\infty |\corpo (\psi,\p,n)|< \infty$ for all Lipschitz
 continuous $\psi$ and $\p$ \cite{StAn07a}.
 To be more specific, the corresponding
SVM is consistent if, e.g., we use the least squares loss and
sequences $\lb_n := n^{-\alpha}$ and $\g_n := n^\beta$, $n \ge 1$, for
fixed $\alpha$ and $\beta$ satisfying $3\alpha > 8d\beta > 0$ and $11
\alpha + 4 \beta < 2$.  
Moreover, if the noise process is also centered then
this SVM actually learns to forecast the next {\em true} state.  
\begin{figure}[h!]
\includegraphics[width=3.6in, height=1.6in, angle=0]{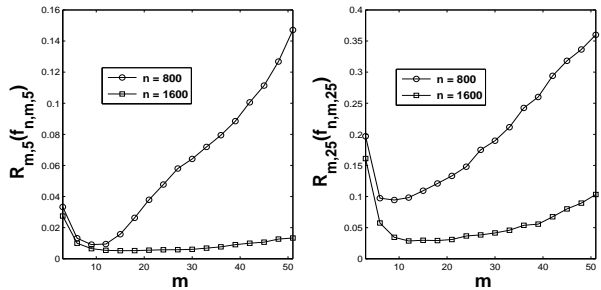}
\caption{{\small Performance of SVM forecasters for $l=5$ (left) and
$l=25$ (right) steps ahead forecasts of the $x$ coordinate of the
Lorenz dynamics with Gaussian noise with $\s=0.05$.  The circles and
squares report, as a function of the memory length $m$, the average 
risk of  25 SVM forecasters built from different samples of size $n = 800$ and $n=
1600$ for each $m$.  }}
\label{svm-forecasts}
\end{figure}

Let us now apply this nonparametric approach to predicting the
evolution of the $x$-coordinate of the Lorenz system described by the
following set of differential equations, $\dot{x} = a (y-x)$, $\dot{y}
= b x - y - xz$, $\dot{z} = xy - c z$, where the parameters are set at
the standard values $a = 10$, $b = 28$, and $c = 8/3$.  The evolution
is sampled at time steps of $\tau = 0.01$ and the observational error
is i.i.d~Gaussain noise. We compute SVM forecasters using memories of
increasing length and for training data of two different sizes $n=800$
and $n=1600$.  Since in estimating the convergence speed of
(\ref{slln-l1}) we are using a loose concentration result, a suitable
regularization parameter and kernel width sequence cannot be chosen
a-priori. Hence, we have adopted a grid search in $(\lb,\g)$ space and
a \mbox{4-fold} cross-validation technique \cite{GyKoKrWa02} to choose
$(\lb_n,\g_n)$ for a given sample size $n$.  Finally, we use
 $(\lb_n,\g_n)$ for an estimate
$f_{n,1,1,\lb_n,\g_n}$ constructed from (\ref{svm-matrix-eq}) using
the whole sample $T_n$ (to simplify notation, we henceforward omit the
dependence of $f_{n,1,1}$ on the regularization parameters
$(\lb_n,\g_n)$).  This approach {\em adapts} the regularization
parameters and the complexity of the SVM forecasters to the amount of
available empirical data. The risk of each forecaster is estimated by
the prediction error over a large test set ($10^5$ input/output pairs)
chosen independent of the training set $T_n$. Note that $\Rx mlf \ge
\sigma^2$, where $\sigma$ is the standard deviation of the noise.

As Fig.~(\ref{svm-forecasts}) shows for $l \in \{5,25\}$, by increasing
the memory of the predictor we obtain forecasters with improved
performance.  However, for a given sample size $n$, there is an {\em
optimal} memory length $m$ and increasing memory length beyond this
value produces poorer forecaster due to their increased complexity for
the available data. Moreover the memory of the best forecaster
increases with sample size from $m \approx 9$ for $n=800$ to $m\approx
18$ for $n=1600$, as shown for $5$-step ahead forecasters, while the
memory increases from $m \approx 9$ for $n=800$ to $m\approx 12$ for
$n=1600$, for $25$-step ahead forecasters.  This reflects the fact
that with increased information we can build more complex and,
therefore, better predictors. Since forecasting further into the
future is a more difficult problem, the performance of the predictor
decreases when we keep the same sample size but attempt to make
$25$-steps ahead predictions.  Furthermore, as the right plot in
Fig.~\ref{svm-forecasts} reveals, for the same sample size $n$ the
memory of the best forecaster reduces in order to accommodate the
increased complexity of the learning task.

\begin{figure}[h!]
\includegraphics[width=3.6in, height=1.6in, angle=0]{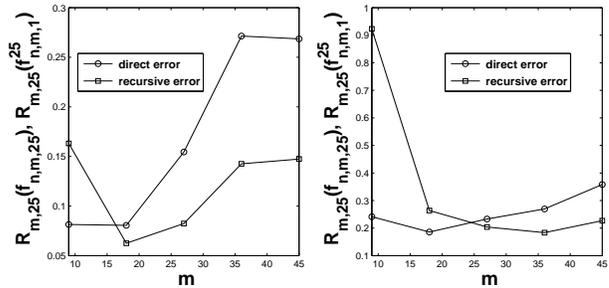}
\caption{{\small Performance of direct and recursive SVM forecasters for
 $l=25$ steps ahead forecasts  of the $x$ coordinate of the
Lorenz dynamics with Gaussian noise with $\s=0.20$ (left) and
$\s=0.05$ (right). The circles and squares report, as a function of the memory length
$m$, the averages forecasting risk of 30 direct and
recursive SVM forecasters built from different samples of size $n =
800$ for each $m$.  }}
\label{rec-dir-forecasts}
\end{figure}
In the limit $ n \rightarrow \infty$  direct forecasts are better than recursive
forecasts because the SVM forecaster approaches the Bayes
forecaster. However, for finite sample sizes this is not always the case.
Indeed, as Fig.~(\ref{rec-dir-forecasts}) shows for $n=800$ 
there is a memory size at which the risk of the direct forecaster,
$f_{n,m,25}$, is larger than that of the recursive, $f_{n,m,1}^{25}$,
$25$-steps ahead forecaster.
Moreover, the crossover depends on the amount of noise and increases
as noise variance increases while keeping the same sample size $n$. By
increasing the sample size though,  the crossover increases
such that in the limit of infinite sample size direct forecasters
always perform better than recursive forecasters, as is expected.

To conclude, we have described the assumptions under which
non-parametric SVM forecasters can consistently learn the optimal
predictors.  For example, for bounded noise, SVM forecasters are
consistent for {\em all} pairs of dynamical systems and observational
noise processes that possess summable correlation sequences. Hence,
the SVM forecasters possess a weak form of universality for a large
class of stochastic processes. This includes systems with smooth
uniformly expanding dynamics or smooth hyperbolic dynamics, systems
perturbed by dynamic noise, as well as "parabolic" or "intermittent"
systems which have a polynomial decay of correlations \cite{Baladi02}.
Remarkably, for some dynamical systems it seems possible to decide on
the summability of the correlation sequence from observations. Indeed,
for piecewise expanding maps rigorous estimates of the asymptotic rate
of decay of correlations for a given function is numerically feasible
\cite{Li00}. 

We also notice that in the presence of observational noise the task of
forecasting is different from the task of modelling the underlying
nonlinear dynamics from data.  Indeed, finding the simplest model
consistent with the observations, which is the ultimate goal of
modelling \cite{KaSc99}, is highly unsuitable for forecasting.  This
is clearly illustrated by noticing the significant increase of
forecasting skill in Fig.~(\ref{svm-forecasts}) when using a memory
length which is much larger than the minimal time delay embedding
necessary to reconstruct a modelling phase space equivalent to the
original state space of the Lorenz system.

Finally, we note that our analysis remains valid for stationary
nonergodic processes.  Since a nonergodic stationary process has an
ergodic decomposition, a realization of the time series falls with
probability one into an invariant event on which the process is ergodic
and stationary. For this reason, the proposed learning algorithm can be applied 
to a time series sequence generated by that event as though it were 
the process universe.


\begin{thebibliography}{15}
\bibitem[{rem()}]{rem1}
\bibinfo{note}{In order to obtain the optimal denoiser we have to minimize a
  modified risk functional, i.e. $\int_{\R^{d}}\int_{\R^{md}} \mnorm{
  F^{m}(x) - f ( z_1,\dots,z_{m} ) }_2^2\, \nu^m (d\e)\mu(dx),$ which
  assumes, unrealistically, exact knowledge of the state, $F^{m}(x)$, that we
  want to estimate by denoising. Nevertheless, by carrying out the minimization
  we obtain Eq.~(\ref{lstep-predictor}) with $l=0$, which is computationally
  feasible when perfect knowledge about the dynamics and the noise is assumed.}
\end{thebibliography}
\end{document}